%% file: 000main.tex
\documentclass[a4paper, 10pt, conference]{IEEEtran}
\usepackage[]{inputenc}
\usepackage{textcomp}

\makeatletter

\title{\LARGE \bf
Non-Negative PARATUCK2 Tensor Decomposition\\
Combined to LSTM Network For Smart Contracts Profiling
}

\usepackage{amsfonts}
\usepackage{amssymb}
\usepackage{amsmath}
\usepackage[mathscr]{euscript}
\usepackage{stmaryrd} 
\usepackage{hyperref} 
\usepackage{graphics} 
\usepackage{bbm} 
\usepackage{dblfloatfix}
\usepackage{blindtext}

\usepackage{tikz}
\usepackage{framed}
\usetikzlibrary{arrows}
\usepackage{standalone}
\usetikzlibrary{decorations.pathreplacing,angles,quotes}

\author{\IEEEauthorblockN{Jeremy Charlier}
\IEEEauthorblockA{SEDAN, SnT\\
University of Luxembourg\\
Luxembourg, Luxembourg\\
jeremy.charlier@uni.lu}
\and
\IEEEauthorblockN{Radu State}
\IEEEauthorblockA{SEDAN, SnT\\
University of Luxembourg\\
Luxembourg, Luxembourg\\
radu.state@uni.lu}
\and
\IEEEauthorblockN{Jean Hilger}
\IEEEauthorblockA{Information Technology\\
BCEE\\
Luxembourg, Luxembourg\\
j.hilger@bcee.lu}
}

\makeatother

\begin{document}

\maketitle

\begin{abstract}

Smart contracts are programs stored and executed on a blockchain.
The Ethereum platform, an open-source blockchain-based platform, has been designed to use these programs offering secured protocols and transaction costs reduction.
The Ethereum Virtual Machine performs smart contracts runs, where the execution of each contract is limited to the amount of gas required to execute the operations described in the code. 
Each gas unit must be paid using Ether, the crypto-currency of the platform. 
Due to smart contracts interactions evolving over time, analyzing the behavior of smart contracts is very challenging. 
We address this challenge in our paper. 
We develop for this purpose an innovative approach based on the non-negative tensor decomposition PARATUCK2 combined with long short-term memory (LSTM) to assess if predictive analysis can forecast smart contracts interactions over time.
To validate our methodology, we report results for two use cases. The main use case is related to analyzing smart contracts and allows shedding some light into the complex interactions among smart contracts. 
In order to show the generality of our method on other use cases, we also report its performance on video on demand recommendation.


\end{abstract}

\section{INTRODUCTION}
\input{Introduction/Intro_gal} %

\section{Related Work}
In this section, we briefly mention the background of DEDICOM TD to its further extension called PARATUCK2. We also underlined related work in smart contracts and LSTM.

\input{RelatedWork/RlWrk_PARATUCK2}

\input{RelatedWork/RlWrk_LSTM} 

\input{RelatedWork/RlWrk_SC_bckgrnd}

\section{PARATUCK2 and LSTM Networks}
In this section, we present in the beginning  mathematical operations involved in PARATUCK2 TD followed by the PARATUCK2 description itself. Finally, LSTM for time prediction is described.

\input{Theory/PRTCK_Description} 

\input{Theory/PRTCK_Decomposition} 

\input{Theory/PRTCK_LSTM}

\section{PARATUCK2-LSTM: a multidisciplinary tensor-RNN based approach }
In this section, we consider two application domains: a small example based on Recommender Systems (RS) for Video On Demand (VOD) such as Netflix and a larger experiment on the Ethereum platform with smart contracts profiling.

The first experiment is performed on a PC with Intel Core i7 CPU and 8 GB of RAM. 
The second experiment is more CPU consuming due to the larger size of the data set and thus uses a more powerful machine with 15 Intel Xeon E5-4650 v4 2.20 Ghz CPU cores and 80 GB of RAM.
We have implemented in the Python programming language the algorithm for non-negative PARATUCK2 decomposition combined with LSTM network. 
\input{Experiments/Rec_Sys/Sim_Movie}

\input{Experiments/SmartContracts/Sim_SmartCntrct}

\section{CONCLUSIONS}
We proposed in this paper  a multi-disciplinary approach leveraging multidimensional linear algebra and neural networks for modeling the complex interactions occurring on a certain type of blockchains.
Our method combines both  of the PARATUCK2 tensor decomposition and LSTM to predict behavior in relation to asymmetric data over time.
The asymmetry is expressed within the tensor decomposition using two sets of latent factors related to two sets of objects. 
We instantiated this model on two use cases.
The first practical case was applied on users and movies as concrete objects. 
The second use case considered sender and receiver contracts of the Ethereum platform.
Our approach allowed to detect in both cases common behaviors over time and was able to predict accurate recommendations in the first case and accurate interactions and exchanges in the second experiment. We validated our results using statistical tests.

Although the method showed good results in terms of accuracy, it currently lacks the required scalability to be used on big data sets. This is due  to  the non-negative ALS update rule which is time and memory consuming.
We plan  to address in future works this issue and develop additional  resolution method to the PARATUCK2 tensor decomposition using other iterative schemes.
Last but not least, the better scalability of the method would help to increase the accuracy of the LSTM network as the training could be performed on longer time period and smaller time step discretization.
We plan to address on particular use-case about fraud detection and detection of suspicious behavior over time.


\section*{ACKNOWLEDGMENT}

The authors would like to thank Beltran Borja Fiz Pontiveros for his help in Ethereum data manipulation. They also thank Jacques Putz and Patric de Waha from Banque et Caisse d'Epargne de l'Etat (BCEE) as one of the strongest support for the research publication.


\bibliographystyle{unsrt}
\bibliography{biblio}

\end{document}

%% file: Introduction/Intro_gal.tex
In the next few months, public institutions and governments would probably start regulating the non-regulated activities of cryptocurrencies such as Bitcoin or Ether as some of the governments already claimed they were investigating cryptocurrencies activities (\cite{Helms2017Bitcoin}, \cite{Woolf2016Bitcoin} and \cite{Rees2017Bitcoin}).
These regulations would probably introduce new sets of rules and would ask for more transparency among the  market players.
As a result,  financial products would probably require key information document (KID) to advise potential investors of the risk of these investments.
Ethereum, with already more than one million accounts, is one of the major platforms for smart contracts relying on Ether cryptocurrency for its existence. 
Still, the platform supports very few documentation about how blockchain players interact and lacks of transparency for non-specialists. 
Modeling smart contracts and predictive analytics is thus essential for future regulation purpose.
Our contributions are twofolds:
\begin{itemize}
\item We describe a PARATUCK2 Tensor Decomposition (TD) for  smart contracts. This decomposition leads to  a a set of latent factors,  where a huge multi-dimensional matrix is decomposed into a less-dimensional  structure.

\item  A second contribution is the prediction of smart contracts activities using LSTM trained on PARATUCK2 TD. The main novelty is that we  predict future activities among a huge quantity of entities by predicting the evolution of a set of latent factors using LSTM.  We used LSTM  since this approach has been shown to learn  from  both long term experience and recent observations. Once the latent factors are predicted, we reconstruct the multidimensional matrices using the predicted latent factors.
\end{itemize}

To outline our approach, we present the related work of tensor decomposition, Long-Short-Term Memory (LSTM) and smart contracts in the first section of the paper. 
Section 2 provides the fundamentals of the chosen tensor decomposition (TD) and how LSTM are applied to the decomposition. 
Section 3 is made from two distinct parts. 
The first part illustrates the concepts of the approach on a small example on Video On Demand (VOD) for recommender systems (RS). 
The second part applies the methodology to the profiling and the predictions of Ethereum smart contracts exchanges among time. 
Finally, we conclude  with pointers to future works.

%% file: RelatedWork/RlWrk_PARATUCK2.tex
\subsection{PARATUCK2 Tensor Decomposition Background}
Harshman, Caroll and Chang have been the first to introduce the concept of the multidimensional tensors and their resolutions with CANDECOMP/PARAFAC (CP) tensor decomposition in \cite{harshman1970foundations} and \cite{carroll1970analysis}.
Since then, CP decomposition has been used in various applications and domains. 
In \cite{giudici2016tensorial}, Giudici and Pecora proposed to use it for the analysis of international bilateral claims.
It underlined existing communities, followed their evolution over time and major event such as the 2008 banking crisis.
In \cite{veganzones2016nonnegative}, the authors applied a novel compression-based CP decomposition to hyperspectral imaging and hyperspectral data.

Furthermore, other decompositions have been introduced afterward such as the DEDICOM decomposition by Harshman in \cite{harshman1978models}.
Unlike the  CP decomposition, it offers the possibility to illustrate asymmetric relationships among data. 
Harshman and Lundy in \cite{harshman1992three} illustrated the concepts of the DEDICOM decomposition by applying the method to international trade data between different nations. 
In \cite{bader2007temporal}, Bader, Harshman, and Kolda proposed a new algorithm for DEDICOM's resolution applied on Enron's email to analyze communities before Enron's crisis. 

For the analysis of interactions between two possibly different sets of objects, Harshman and Lundy introduced the PARATUCK2 decomposition in \cite{harshman1996uniqueness}.
The algorithm has been applied by Bro in \cite{bro1998multi} to flow injection analysis system with a constraint on pH-gradients. 
In the recent years, the decomposition has been also applied in communication networks. 
In \cite{kibangou2007blind}, Kibangou and Favier considered a special design of an input signal in communication channels to show that the output signal could be represented using PARATUCK2 representation. 
The authors in \cite{rui2016novel} proposed a novel algorithm for its resolution applied to Multiple Input Multiple Output (MIMO) relay communication for faster convergence.
However, the drawback of PARATUCK2 is that the decomposition results in a complex factorization leading to non-trivial analysis which could explain why it has been less commonly applied among scientific community.

%% file: RelatedWork/RlWrk_LSTM.tex
\subsection{LSTM as one of the state-of-the-art neural network}
The problem of the vanishing gradients in  Recurrent Neural Networks (RNN) was solved in an elegant way in  \cite{hochreiter1997long} by Hochreiter and Schmidhuber through the LSTM architecture, which subsequent found multiple applications in a broad area of applications that range from speech recognition and hand writing to social media analysis.
For a comprehensive overview on the application domains, the reader is refered to  \cite{greff2017lstm},  where the authors describe  eight of the most common LSTM variants on different domains.

The authors in \cite{alahi2016social} illustrated the efficiency of LSTM in social science by training a  model on human trajectories in crowded environment. 
Once trained, the authors researched if the network was able to predict future large movement of persons and to avoid collisions between humans.
The modeling and the predictions of pure time series have been showed in \cite{malhotra2015long} with the study of the resilience of LSTM.
In their results, the authors highlighted for various domains such as space shuttle or power demand that a trained network on non-anomalous data could identify anomalies for unknown length of time series. 

Based on these previous results, illustrating the efficiency and the scalability of LSTM  for time series forecasting, we consider in this paper the extent to which the behaviors of smart contracts can  be identified and predicted using LSTM.

%% file: RelatedWork/RlWrk_SC_bckgrnd.tex
\subsection{Emergence Of Smart Contracts On Distributed Ledger}
Introduced in 1994 by Nick Szabo as "\textit{a computerized transaction protocol that executes the terms of a contract}",  a smart contract  allows public execution across a distributed networks of nodes through a democratic organization. This is done without requiring a central gatekeeper.
We witness in the present an evolution towards real world application domains and business by major industry players. 
As such, in \cite{frantz2016institutions}, Frantz and Nowostawski proposed a modeling approach of smart contracts that could transpose human-readable contract into computational equivalents for the contracts execution on the blockchain.
The goal of the mechanism is to allow the specification and the interpretation of the smart contracts to larger audience than just blockchain specialists.
In \cite{christidis2016blockchains}, Christidis and Devetsikiotis tried to find some applications of the blockchain into Internet of Things (IoT). 
Although most of the research is optimistic about broader applications of smart contracts, others raised concerns about security. 
More specifically, in \cite{atzei2017survey}, Atzei, Bartoletti and Cimoli tested the smart contracts implementation in Ethereum against attacks with the goal of stealing or tampering assets transferred through smart contracts. 
In their experiments, they showed how a series of attack could lead to a money robbery or to a denial of service attack against the Ethereum blockchain.
The cited papers illustrate a common underlying approach: most of the works either extend the use of the smart contracts or highlight the security concerns due to their usage. 

In our approach, we propose an innovative application of the PARATUCK2 tensor decomposition for the analysis of smart contracts exchanges. 
We also combine in a novel approach the PARATUCK2 tensor decomposition with LSTM inherited from neural networks. We show that our method can accurately predict future probable exchanges between smart contracts.

%% file: Theory/PRTCK_Description.tex
\subsection{Tensor Description}
\textbf{Notation} Terminology in this paper follows  the one described by Kolda and Bader in \cite{kolda2009tensor} and commonly used by other publications. Scalars are denoted by lower case letters, \textit{a}. Vectors and matrices are described by boldface lowercase letters and boldface capital letters, respectively \textbf{a} and \textbf{A}. High order tensors are represented using Euler script notation such as $\mathscr{X}$.

The transpose matrix of $A\in \mathbb{R}^{I\times J}$ is denoted by $A^T$ and results in a matrix of size $\mathbb{R}^{J\times I}$.

The Moore-Penrose inverse of a matrix $A\in \mathbb{R}^{I\times J}$ is denoted by $A^\dag$ and results in matrix of size $\mathbb{R}^{J\times I}$.

\textbf{Tensor Definition} $\mathscr{X}$ is called a \textit{n}-way tensor if  $\mathscr{X}$ is a \textit{n}-th multidimensional array and can be expressed as $\mathscr{X}\in \mathbb{R}^{I_1 \times I_2 \times I_3 \times ... \times I_n}$.

\textbf{Tensor Operations} The square root of the sum of all tensor entries squared of the tensor $\mathscr{X}$ defines its norm.
\begin{equation} \label{eq::norm}
||\mathscr{X}||=\sqrt{\sum_{j=1}^{I_1}\sum_{j=2}^{I_2}...\sum_{j=n}^{I_n}x_{j_1, j_2, ..., j_n}^2}
\end{equation}
The rank-\textit{R} of a tensor $\mathscr{X}\in\mathbb{R}^{I_1\times I_2\times ...\times I_N}$ is the number of linear components that could fit $\mathscr{X}$ exactly such that
\begin{equation} \label{eq::rank}
\mathscr{X}=\sum_{r=1}^R \textbf{a}_r^{(1)} \circ \textbf{a}_r^{(2)} \circ ... \circ \textbf{a}_r^{(N)}
\end{equation}
with the symbol $\circ$ representing the vector outer product.

The Kronecker product between two matrices \textbf{A}$\in\mathbb{R}^{I\times J}$ and \textbf{B}$\in\mathbb{R}^{K\times L}$, denoted by \textbf{A}$\otimes$\textbf{B}, results in a matrix \textbf{C}$\in\mathbb{R}^{IK\times KL}$.
\begin{equation} \label{eq::kron}
\textbf{C}=\textbf{A}\otimes\textbf{B}=
\begin{bmatrix}
 a_{11}\textbf{B}& a_{12}\textbf{B} & \cdots & a_{1J}\textbf{B}\\ 
 a_{21}\textbf{B}& a_{22}\textbf{B} & \cdots & a_{2J}\textbf{B}\\
 \vdots & \vdots & \ddots & \vdots \\ 
 a_{I1}\textbf{B}& a_{I2}\textbf{B} & \cdots & a_{IJ}\textbf{B}
\end{bmatrix}
\end{equation}

The Khatri-Rao product between two matrices \textbf{A}$\in\mathbb{R}^{I\times K}$ and \textbf{B}$\in\mathbb{R}^{J\times K}$, denoted by \textbf{A}$\odot$\textbf{B}, results in a matrix \textbf{C} of size $\mathbb{R}^{IJ\times K}$. It is the column-wise Kronecker product.
\begin{equation} \label{eq::kr}
\textbf{C}=\textbf{A}\odot\textbf{B}=
[\textbf{a}_1\otimes \textbf{b}_1 \quad \textbf{a}_2\otimes \textbf{b}_2 \quad \cdots \quad \textbf{a}_K\otimes \textbf{b}_K]
\end{equation}

%% file: Theory/PRTCK_Decomposition.tex
\subsection{PARATUCK2 Tensor Decomposition}
\begin{figure}[b]
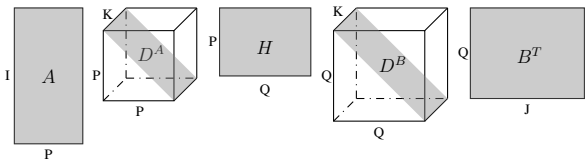

\begin{center}
\includestandalone{tikz_fig/tikz_PARATUCK2}
\caption{PARATUCK2 decomposition of a three-way tensor with dimension notations}
\label{fig::PARATUCK2}
\end{center}
\end{figure}

In our approach, we use the PARATUCK2 decomposition introduced by Harshman and Lundy in \cite{harshman1996uniqueness} . 
This decomposition is well suited for the analysis of intrinsically asymmetric relationships between two different sets of objects. 
It represents a tensor $\mathscr{X}\in\mathbb{R}^{I\times J\times K}$ as a product of matrices and tensors.

\begin{equation} \label{eq::paratuck2}
\textbf{X}_k = \textbf{A}\textbf{D}^A_k\textbf{H}\textbf{D}^B_k\textbf{B}^T \quad \textrm{with} \quad k=\left\lbrace 1, ..., K \right\rbrace
\end{equation}
The matrices $\textbf{A}$, $\textbf{H}$ and $\textbf{B}$ are matrices of size $\mathbb{R}^{I\times P}$, $\mathbb{R}^{P \times Q}$ and $\mathbb{R}^{J\times Q}$.
The matrices $\textbf{D}^A_k\in \mathbb{R}^{P\times P}$ and $\textbf{D}^B_k\in \mathbb{R}^{Q\times Q}\:\forall k\in\left\lbrace 1, ...,K \right\rbrace$ are the slices of the tensors $\mathscr{D}^A\in \mathbb{R}^{P\times P \times K}$ and $\mathscr{D}^B\in \mathbb{R}^{Q\times Q \times K}$. The latent factors $P$ and $Q$ are related to the rank of each object set as illustrated in figure \ref{fig::PARATUCK2}.
The columns of the matrices $\textbf{A}$ and $\textbf{B}$ represent the latent factors $P$ and $Q$, respectively. The matrix $\textbf{H}$ describes the asymmetry between the $P$ latent factors and the $Q$ latent factors. Finally, the tensors $\mathscr{D}^A$ and $\mathscr{D}^B$ measures the evolution of the latent factors regarding the third dimension. 

To achieve the computation of the PARATUCK2 decomposition, the following minimization equation has to be solved
\begin{equation} \label{eq::minimztn}
\min_{\mathscr{\hat{X}}} ||\mathscr{X}-\mathscr{\hat{X}}||
\end{equation}
with $\mathscr{\hat{X}}$ the approximate tensor described by the decomposition and $\mathscr{X}$ the original tensor.

To solve equation \ref{eq::minimztn}, the Alternating Least  Squares (ALS) method is used as presented by Bro in \cite{bro1998multi}. All of the matrices and the tensors are updated iteratively. To simplify the resolution explanation, we consider one level \textit{k} of \textit{K}, the third dimension of the tensor.

To update $\textbf{A}$, equation \ref{eq::paratuck2} is rearranged such that
\begin{equation} \label{eq::upd_A}
\textbf{X}_k=\textbf{A}\textbf{F}_k 
\quad \textrm{with} \quad
\textbf{F}_k=\textbf{D}_k^A \textbf{H} \textbf{D}_k^B \textbf{B}^T
\end{equation}
The simultaneous least square solution for all \textit{k} leads to 
\begin{equation}
\textbf{A} = \textbf{X}(\textbf{F}^\dag)^T
\quad \textrm{with} \quad
\begin{cases}
\textbf{X} = & \left[ \textbf{X}_1 \: \textbf{X}_2 \cdots \textbf{X}_k \right] \\ 
\textbf{F} = & \left[ \textbf{F}_1 \: \textbf{F}_2 \cdots \textbf{F}_k \right] \\ 
\end{cases}
\end{equation}

To update $\mathscr{D}^A$, equation \ref{eq::paratuck2} is rearranged such that 
\begin{equation} \label{eq::upd_DA}
\textbf{X}_k=\textbf{A} \textbf{D}_k^A \textbf{F}_k^T
\quad \textrm{with} \quad
\textbf{F}_k=\textbf{B} \textbf{D}_k^B \textbf{H}^T
\end{equation}
The matrix $\textbf{D}_k^A$ is a diagonal matrix which lead to the below resolution.
\begin{equation}
\textbf{D}_{(k,:)}^A=\left[ (\textbf{F}_k \odot \textbf{A}) \textbf{x}_k \right]^T
\quad \textrm{with} \quad 
\textbf{x}_k = \textrm{vec}(\textbf{X}_k)
\end{equation}
The notation $(k,:)$ represents the \textit{k}th row of $\textbf{D}_{(k,:)}^A$.

To update $\textbf{H}$, equation \ref{eq::paratuck2} is rearranged such that 
\begin{equation} \label{eq::upd_H}
\textbf{x}_k = (\textbf{B}\textbf{D}_k^B \otimes \textbf{A}\textbf{D}_k^A) \textbf{h}
\quad \textrm{with} \quad
\begin{cases}
\textbf{x}_k = & \textrm{vec}(\textbf{X}_k) \\ 
\textbf{h} = & \textrm{vec}(\textbf{H}) \\ 
\end{cases}
\end{equation}
which brings the solution 
\begin{equation}
\textbf{h} = \textbf{Z}^\dag \textbf{x}
\quad \textrm{with} \quad
\textbf{Z} = \begin{pmatrix}
\textbf{B}\textbf{D}_1^B \otimes \textbf{A}\textbf{D}_1^A \\ 
\textbf{B}\textbf{D}_2^B \otimes \textbf{A}\textbf{D}_2^A \\ 
\vdots \\
\textbf{B}\textbf{D}_k^B \otimes \textbf{A}\textbf{D}_k^A
\end{pmatrix}
\end{equation}

To update $\textbf{B}$ and $\mathscr{D}^B$, the  methodology presented for the update of $\textbf{A}$ and $\mathscr{D}^A$ is applied respectively.

In the experiments, we use the non-negative PARATUCK2 decomposition leveraging the  non-negative matrix factorization presented by Lee and Seung in \cite{lee1999learning}. 
The matrices $\textbf{A}$, $\textbf{B}$ and $\textbf{H}$, and the tensors $\mathscr{D}^A$ and $\mathscr{D}^B$ are computed according to the following multiplicative update rule.
\begin{equation} \label{eq::als_upd}
\begin{cases}
a_{ip}\leftarrow a_{ip} 
\dfrac{\left[ \textbf{X} \textbf{F}^T\right]_{ip}}
{\left[ \textbf{A}(\textbf{FF}^T) \right]_{ip}} 
\: &, \:
\textbf{F}=\mathscr{D}^A \textbf{H}\mathscr{D}^B \textbf{B}^T\\
d^a_{pp}\leftarrow d^a_{pp}
\dfrac{\left[ \textbf{Z}^T \textbf{x} \right]_{pp}}
{\left[ \mathscr{D}^A(\textbf{ZZ}^T) \right]_{pp}} 
\: &, \:
\textbf{Z}=(\textbf{B}\mathscr{D}^B \textbf{H}^T) \odot\textbf{A} \\
h_{pq}\leftarrow h_{pq}
\dfrac{\left[ \textbf{Z}^T \textbf{x} \right]_{pq}}
{\left[ \textbf{H}(\textbf{ZZ}^T) \right]_{pq}} 
\: &, \:
\textbf{Z}=\textbf{B} \mathscr{D}^B \otimes \textbf{A} \mathscr{D}^A \\
d^b_{qq}\leftarrow d^b_{qq}
\dfrac{\left[ \textbf{x} \textbf{Z} \right]_{qq}}
{\left[ \mathscr{D}^B (\textbf{Z}^T\textbf{Z})\right]_{qq}}
\: &, \:
\textbf{Z}=\textbf{B} \odot (\textbf{H}^T \mathscr{D}^A \textbf{A}^T)^T\\
b_{qj}\leftarrow b_{qj} 
\dfrac{\left[ \textbf{X}^T \textbf{F}^T \right]_{qj}}
{\left[ \textbf{B}(\textbf{F} \textbf{F}^T) \right]_{qj}} 
\: &, \:
\textbf{F}= (\textbf{A} \mathscr{D}^A \textbf{H}\mathscr{D}^B)^T\\
\end{cases}
\end{equation}
with 
\begin{equation}
\begin{cases}
\textbf{X} &= \left[ \textbf{X}_1 \: \textbf{X}_2 \cdots \textbf{X}_k \right] \\ 
\textbf{x} &= \textrm{vec}(\mathscr{X}) 
\end{cases}
\end{equation}

The multiplicative update rule helps to better calibration of LTSM that uses the elements of the tensor decomposition as a starting point.

%% file: tikz_fig/tikz_PARATUCK2.tex
\begin{tikzpicture}[scale=0.6, transform shape]
\draw  (-4.5,2) rectangle (-3,-1);
\fill [gray, opacity=.4]  (-4.5,2) rectangle (-3,-1);
\draw (-3.75,0.75) node[below]{\large{$A$}};

\draw (2.5,-0.5) rectangle (4.5,1.5) node (v1) {};
\draw (4.5,1.5) -- (5,2);
\draw (2.5,1.5) -- (3,2);
\draw (4.5,-0.5) -- (5,0);
\draw (3,2) -- (5,2);
\draw (5,0) -- (5,2);
\draw [dashdotted] (3,0) -- (5,0);
\draw [dashdotted] (3,0) -- (3,2);
\draw [dashdotted] (2.5,-0.5) -- (3,0);
\fill [gray, opacity=.4] (4.5,-0.5) -- (5,0) -- (3,2) -- (2.5,1.5)  -- cycle;
\draw (-1.5,1.3) node[below]{\large{$D^A$}};

\draw (0,2) rectangle (2,0.5);
\fill [gray, opacity=.4] (0,2) rectangle (2,0.5);
\draw (1,1.4) node[below]{\large{$H$}};

\draw (-2.55,-0.05) rectangle (-1,1.5) node (v1) {};
\draw (-1,1.5) -- (-0.5,2);
\draw (-2.55,1.5) -- (-2.05,2);
\draw (-1,-0.05) -- (-0.5,0.45);
\draw (-2.05,2) -- (-0.5,2);
\draw (-0.5,0.45) -- (-0.5,2);
\draw [dashdotted] (-2.05,0.45) -- (-0.5,0.45);
\draw [dashdotted] (-2.05,0.45) -- (-2.05,2);
\draw [dashdotted] (-2.55,-0.05) -- (-2.05,0.45);
\fill [gray, opacity=.4] (-1,-0.05) -- (-0.5,0.45) -- (-2.05,2.0) -- (-2.55,1.5)  -- cycle;
\draw (3.8,1.05) node[below]{\large{$D^B$}};

\draw  (5.5,2) rectangle (8,0);
\fill [gray, opacity=.4] (5.5,2) rectangle (8,0);
\draw (6.8,1.25) node[below]{\large{$B^T$}};

\draw (-4.65,0.75) node[below]{\small{I}};
\draw (-3.75,-1) node[below]{\small{P}};
\draw (-1.75,-0.05) node[below]{\small{P}};
\draw (-2.7,0.75) node[below]{\small{P}};
\draw (-2.45,2.1) node[below]{\small{K}};
\draw (2.6,2.15) node[below]{\small{K}};
\draw (-0.15,1.5) node[below]{\small{P}};
\draw (1,0.45) node[below]{\small{Q}};
\draw (2.35,0.75) node[below]{\small{Q}};
\draw (3.5,-0.5) node[below]{\small{Q}};
\draw (5.35,1.25) node[below]{\small{Q}};
\draw (6.75,0) node[below]{\small{J}};
\end{tikzpicture}

%% file: Theory/PRTCK_LSTM.tex
\subsection{LSTM applied to PARATUCK2}
Our contribution contains besides an application of  the non-negative PARATUCK2 scheme to smart contracts, also a computational step based on LSTM. 
LSTM has been introduced in \cite{hochreiter1997long} by Hochreiter and Schmidhuber with the goal of overcoming the short-ends of RNN. 
As described by the authors in \cite{cho2014learning}, a RNN is a neural network with a hidden state, $\textbf{h}$, and an optional output $y$. 
It operates on an event sequence $\textbf{x}$. 
RNN suffers from the vanishing gradient which leads to the loss of long-term memory. 
The loss of the long-term memory was solved by the LSTM architecture.

Following the notation of Sak, Senor and Beaufays
in \cite{sak2014long}, LSTM contains memory blocks in the
recurrent hidden layer. 
Each memory block is connected to an input gate and an output gate. Similarly to RNN, the input gate plays the role of the input activation of the memory cells. 
The output gate is in charge of the  flow of cell activations into the rest of the network. 
In addition, a forget gate is added to the memory block since Gers, Cummins and Schmidhuber presented it in \cite{gers1999learning}.
The forget gate allows the reset of the cell's memory depending on the information received through the input gate. 
If we  consider the input sequence denoted by $x$ such as $x=(x_1, \cdots, x_T)$, the output sequence denoted by $y$ such as $y=(y_1, \cdots, y_T)$ for a sequence of events from $t=1$ to $t=T$. 
The mapping between $x$ and $y$ for all network unit activations within LSTM is described by a set of equations. 
The activation of the input gate is denoted by $i_t$, the candidate value for the states of the memory cells by $\tilde{C}_t$, the activation of the memory cells’ forget gates by $f_t$, the memory cells’ new state by $C_t$, the value of their output gates by $o_t$ and the outputs of the output gates by $h_t$.

\begin{equation} \label{eq::LSTM_ntwrk}
\begin{cases}
i_t &= \sigma(W_{i}x_t+U_{i}h_{t-1}+b_i)\\
\tilde{C}_t &= tanh(W_c x_t+U_c h_{t-1}+b_c)\\
f_t &= \sigma(W_{f}x_t+U_fh_{t-1}+b_f)\\
C_t &= i_t\ast \tilde{C}_t+f_t\ast C_{t-1}\\
o_t &= \sigma(W_{o}x_t+U_{o}h_{t-1}+V_{o}C_{t}+b_o)\\
h_t &= o_t \ast tanh(C_t)
\end{cases}
\end{equation}

In the set of equations \ref{eq::LSTM_ntwrk} at time $t$, $x_t$ stands for the memory cell layer, $W_k$ and $U_k$ with $k=\left\lbrace i,c,f,o \right\rbrace$ for the weight matrices and $b_k$ for the bias vectors.
In the model used for the experiments, the activation of a cell{'}s output gate is independent of the memory cell{'}s state $C_t$ such that $V_0=0$. The main advantage by fixing $V_0=0$ is the ability to perform faster computation, especially on large datasets.

\begin{figure}[b!]
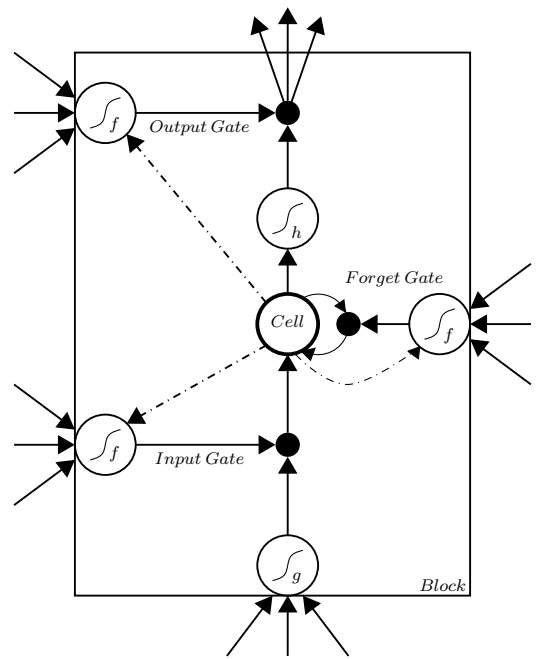

\begin{center}
\includestandalone{tikz_fig/tikz_LSTM_Ntwrk}
\caption{Overview of a LSTM memory cell. In our model, the activation functions \textit{g} and \textit{h} are described by \textit{tanh} and \textit{f} is the forget gate.}
\label{fig::LSTM}
\end{center}
\end{figure}

\begin{figure}[t!]
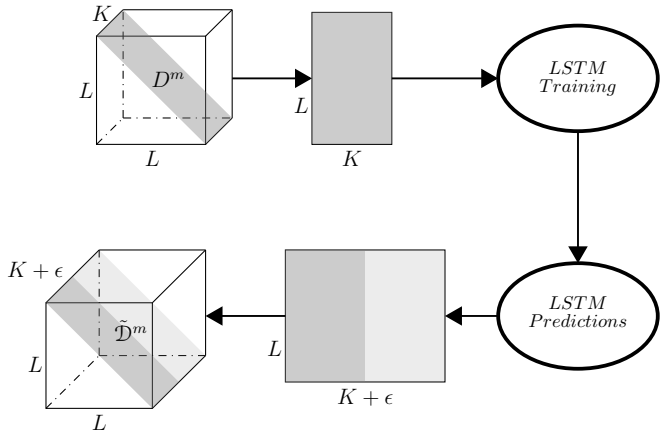

\begin{center}
\includestandalone{tikz_fig/tikz_LSTM_trng}
\caption{Overview of LSTM training and predictions on the tensor $\mathscr{D}^m\in \mathbb{R}^{L\times L \times K}$ with $m=\lbrace A, B \rbrace$ and $L=\lbrace P, Q \rbrace$}
\label{fig::LSTM_prdct}
\end{center}
\end{figure}

With regards to figure \ref{fig::PARATUCK2}, the tensors $\mathscr{D}^A$ and $\mathscr{D}^B$ collects data about the tensor factorization related to the third dimension, which is very often the time.
It means that the evolution of each groups, or clusters, characterized by the latent factors $P$ and $Q$ of the TD contained in the tensors $\mathscr{D}^A$ and $\mathscr{D}^B$ can be modeled using LSTM. 
More precisely, LSTM is calibrated on the historical data of the tensors $\mathscr{D}^A$ and $\mathscr{D}^B$ to predict afterwards the future evolution of each P and Q groups contained in the tensors $\mathscr{D}^A$ and $\mathscr{D}^B$ as illustrated in figure \ref{fig::LSTM_prdct}.

Only the diagonals of the tensors $\mathscr{D}^m$ with $m=\lbrace A,B \rbrace$ contain numbers which means that the tensors $\mathscr{D}^m\in \mathbb{R}^{L\times L \times K}$ can be reduced to a matrix, $\textbf{E}$, of size $\mathbb{R}^{L\times K}$. 
The notation $L=\lbrace P,$ $Q \rbrace$ denotes the latent factors of the PARATUCK2 TD.
Data contained in $\textbf{E}$ is then used to train LSTM before performing the predictions on an interval $\epsilon$ related to the third dimension $K$.
The resulting matrix of size $\mathbb{R}^{L\times (K+\epsilon)}$ gathers the historical data of each latent component $L$ as well as the predicted values.
A new tensor denoted by $\tilde{\mathscr{D}}^m$ of size $\mathbb{R}^{L\times L\times (K+\epsilon)}$ is then built. 
The methodology is applied on both tensors $\mathscr{D}^A$ and $\mathscr{D}^B$ for the same $\epsilon$ which leads afterward to a PARATUCK2 tensor decomposition linked to historical data as well as predicted data.

%% file: tikz_fig/tikz_LSTM_Ntwrk.tex
\begin{tikzpicture}[scale=0.80, transform shape]
\draw  [line width=0.25mm] (0,-5) circle (0.5);
\fill [opacity=1, line width=0.25mm] (0,-3) circle (0.20);
\draw [draw=black, line width=0.25mm, -triangle 60] (0,-4.5) -- (0,-3.2);

\draw  [line width=0.25mm] (-3,-3) circle (0.5);
\draw [draw=black, line width=0.25mm, -triangle 60] (-2.5,-3) -- (-0.2,-3);

\draw [draw=black, line width=0.25mm, -triangle 60] (-0,-2.8) -- (-0,-1.5);

\draw  [line width=0.5mm] (0,-1.0) node (v2) {} circle (0.5);
\fill [opacity=1, line width=0.25mm] (1,-1) node (v1) {} circle (0.20);
\draw  [line width=0.25mm] (2.5,-1) circle (0.5);

\draw [draw=black, line width=0.25mm, -triangle 60] (2,-1) -- (1.20,-1);

\draw (0.25,-0.57) arc (120.0:20:0.5);
\draw (0.25,-1.44) arc (-120.0:20:0.5);
\draw[fill=black] (0.775,-0.7) -- (0.95,-0.8) -- (0.925,-0.6)-- (0.775,-0.7);
\draw[fill=black] (0.225,-1.45) -- (0.45,-1.4) -- (0.375,-1.6)-- (0.225,-1.45);

\draw [draw=black, line width=0.25mm, -triangle 60] (0,-0.5) -- (0,0.25);
\draw  [line width=0.25mm] (0,0.75) circle (0.5);
\draw [draw=black, line width=0.25mm, -triangle 60] (0,1.25) -- (0,2.3);

\fill [opacity=1, line width=0.25mm] (0,2.5) circle (0.20);
\draw  [line width=0.25mm] (-3,2.5) circle (0.5);
\draw [draw=black, line width=0.25mm, -triangle 60] (-2.5,2.5) -- (-0.2,2.5);

\draw [draw=black, line width=0.25mm, -triangle 60] (0,2.7) -- (0,4.25);
\draw [draw=black, line width=0.25mm, -triangle 60] (0,2.5) -- (-0.55,4.1);
\draw [draw=black, line width=0.25mm, -triangle 60] (0,2.5) -- (0.55,4.1);

\draw [draw=black, line width=0.25mm, -triangle 60] (-4.5,2.5) -- (-3.5,2.5);
\draw [draw=black, line width=0.25mm, -triangle 60] (-4.5,3.5) -- (-3.5,2.75);
\draw [draw=black, line width=0.25mm, -triangle 60] (-4.5,1.5) -- (-3.5,2.25);

\draw [draw=black, line width=0.25mm, -triangle 60] (-4.5,-3) -- (-3.5,-3);
\draw [draw=black, line width=0.25mm, -triangle 60] (-4.5,-2) -- (-3.5,-2.75);
\draw [draw=black, line width=0.25mm, -triangle 60] (-4.5,-4) -- (-3.5,-3.25);

\draw [draw=black, line width=0.25mm, -triangle 60] (0,-6.5) -- (0,-5.5);
\draw [draw=black, line width=0.25mm, -triangle 60] (-1,-6.5) -- (-0.25,-5.5);
\draw [draw=black, line width=0.25mm, -triangle 60] (1,-6.5) -- (0.25,-5.5);

\draw [draw=black, line width=0.25mm, -triangle 60] (4,-1) -- (3,-1);
\draw [draw=black, line width=0.25mm, -triangle 60] (4,0) -- (3,-0.75);
\draw [draw=black, line width=0.25mm, -triangle 60] (4,-2) -- (3,-1.25);

\draw [dashdotted, draw=black, line width=0.25mm, -triangle 60] (-0.35,-0.65) -- (-2.65,2.15);
\draw [dashdotted, draw=black, line width=0.25mm, -triangle 60] (-0.35,-1.35) -- (-2.65,-2.65);

\draw [line width=0.25mm] (-3.5,3.5) rectangle (3,-5.5);
\draw (2.55,-5.1) node[below]{\footnotesize{$Block$}};
\draw (-1.45,-3) node[below]{\footnotesize{$Input \: Gate$}};
\draw (-1.45,2.5) node[below]{\footnotesize{$Output \: Gate$}};
\draw (1.75,0) node[below]{\footnotesize{$Forget \: Gate$}};
\draw (0,-0.725) node[below]{\footnotesize{$Cell$}};

\draw  [dashdotted] plot[smooth, tension=.7] coordinates {(0.125,-1.5) (1,-2) (2.15,-1.4)};
\draw[fill=black] (1.975,-1.45) -- (2.15,-1.575) -- (2.17,-1.377)-- (1.975,-1.45);

\draw  plot[smooth, tension=.7] coordinates {(-3.25,2.25) (-3.05,2.35) (-2.95,2.65) (-2.75,2.75)};
\draw  plot[smooth, tension=.7] coordinates {(-0.25,0.5) (-0.05,0.6) (0.05,0.9) (0.25,1)};
\draw  plot[smooth, tension=.7] coordinates {(2.25,-1.25) (2.45,-1.15) (2.55,-0.85) (2.75,-0.75)};
\draw  plot[smooth, tension=.7] coordinates {(-3.25,-3.25) (-3.05,-3.15) (-2.95,-2.85) (-2.75,-2.75)};
\draw  plot[smooth, tension=.7] coordinates {(-0.25,-5.25) (-0.05,-5.15) (0.05,-4.85) (0.25,-4.75)};

\draw (0.15,-5.0) node[below]{\footnotesize{$g$}};
\draw (0.15,0.75) node[below]{\footnotesize{$h$}};
\draw (-2.85,-2.9) node[below]{\footnotesize{$f$}};
\draw (-2.85,2.55) node[below]{\footnotesize{$f$}};
\draw (2.65,-0.95) node[below]{\footnotesize{$f$}};
\end{tikzpicture}

%% file: tikz_fig/tikz_LSTM_trng.tex
\begin{tikzpicture}[scale=0.7, transform shape]
\draw (-1.05,0) rectangle (1,2.05) node (v1) {};
\draw (1,2.05) -- (1.5,2.55);
\draw (-1.05,2.05) -- (-0.55,2.55);
\draw (1,0) -- (1.5,0.5);
\draw (-0.55,2.55) -- (1.5,2.55);
\draw (1.5,0.5) -- (1.5,2.55);
\draw [dashdotted] (-0.55,0.5) -- (1.5,0.5);
\draw [dashdotted] (-0.55,0.5) -- (-0.55,2.55);
\draw [dashdotted] (-1.05,0) -- (-0.55,0.5);
\fill [gray, opacity=.4] (1,0) -- (1.5,0.5) -- (-0.55,2.55) -- (-1.05,2.05)  -- cycle;
\draw (0.3,1.5) node[below]{\large{$D^m$}};

\draw [draw=black, line width=0.25mm, -triangle 60] (1.5,1.25) -- (3,1.25);

\draw  (3,2.5) rectangle (4.5,0);
\fill [gray, opacity=.4]  (3,2.5) rectangle (4.5,0);

\draw [draw=black, line width=0.25mm, -triangle 60] (4.5,1.25) -- (6.5,1.25);

\draw  [line width=0.5mm] (8,1.25) ellipse (1.5 and 1.0);
\draw (8,1.70) node[below]{{$LSTM$}};
\draw (8,1.35) node[below]{{$Training$}};

\draw  [line width=0.5mm] (8,-3.25) ellipse (1.5 and 1.0);
\draw (8,-2.75) node[below]{{$LSTM$}};
\draw (8,-3.10) node[below]{{$Predictions$}};

\draw [draw=black, line width=0.25mm, -triangle 60] (8,0.25) -- (8,-2.25);
\draw [draw=black, line width=0.25mm, -triangle 60] (6.5,-3.25) -- (5.5,-3.25);

\draw  (2.5,-4.5) rectangle (5.5,-2);
\fill [gray, opacity=.4]  (2.5,-4.5) rectangle (4,-2);
\fill [gray, opacity=.15]  (4,-4.5) rectangle (5.5,-2);

\draw (-2,-5) rectangle (0,-3) node (v1) {};
\draw (0,-3) -- (1,-2);
\draw (-2,-3) -- (-1,-2);
\draw (0,-5) -- (1,-4);
\draw (-1,-2) -- (1,-2);
\draw (1,-4) -- (1,-2);
\draw [dashdotted] (-1,-4) -- (1,-4);
\draw [dashdotted] (-1,-4) -- (-1,-2);
\draw [dashdotted] (-2,-5) -- (-1,-4);
\fill [gray, opacity=.4] (0,-5) -- (0.5,-4.5) -- (-1.5,-2.5) -- (-2,-3)  -- cycle;
\fill [gray, opacity=.15] (0.5,-4.5) -- (1,-4) -- (-1,-2) -- (-1.5,-2.5)  -- cycle;
\draw (-0.4,-3.2) node[below]{\large{$\tilde{\mathscr{D}}^m$}};

\draw [draw=black, line width=0.25mm, -triangle 60] (2.5,-3.25) -- (1,-3.25);

\draw (2.8,1) node[below]{\large{$L$}};
\draw (0,0) node[below]{\large{$L$}};
\draw (-1.25,1.3) node[below]{\large{$L$}};
\draw (2.3,-3.6) node[below]{\large{$L$}};
\draw (-1,-5) node[below]{\large{$L$}};
\draw (-2.2,-3.9) node[below]{\large{$L$}};

\draw (3.75,0) node[below]{\large{$K$}};
\draw (-0.95,2.75) node[below]{\large{$K$}};
\draw (4,-4.5) node[below]{\large{$K+\epsilon$}};
\draw (-2.2,-2.1) node[below]{\large{$K+\epsilon$}};
\end{tikzpicture}

%% file: Experiments/Rec_Sys/Sim_Movie.tex
\subsection{A first approach of PARATUCK2-LSTM concepts using recommender systems}
In this first example, the objective is to illustrate the main concepts of the proposed method through a simple application.  We aim at familiarizing the reader with the understanding of latent factors, matrices and tensors involved in the PARATUCK2 tensor decomposition and how these  are combined with LSTM.
Hereinafter, the ability of the model to provide personalized movie recommendations is addressed.

A RS is a methodology that analyzes a large volume of dynamically generated information to provide users with personalized content and services (\cite{isinkaye2015recommendation}).

The data for this experiment was downloaded from the Movielens data base provided by the University of Minnesota which contains more than 20 millions ratings for more than 135,000 users and 27,000 movies.
Since our aim is to provide a small but illustrative  example, the dataset used for the experiments has been shortened to 100 users and 125 movies randomly selected from 10 April 2014 to 30 March 2015.
The data set is divided in two data sets, one for training from 10 April 2014 to 10 November 2014, and one for simulation and validation that ranges  from 11 November 2014 to 30 March 2015.
The second dataset allows to cross-validate the prediction simulation with the events that effectively happened and assess the accuracy of the simulation.

Two three-way tensors, denoted by $\mathscr{X}\in\mathbb{R}^{I\times J \times K}$, are used for the experiments.
The first dimension of $\mathscr{X}$, denoted by $I$, represents the users, the second dimension, $J$, models the movies and the third dimension, $K$, represents the time. 
Each tensor is generated by counting how many times a user watched a specific movie within a particular time period.
The first tensor of size $\mathbb{R}^{I\times J \times K}$ is completed with the first dataset. 
After the tensor decomposition is performed, LSTM is trained between 10 April 2014 and 10 November 2014. We use the LSTM to   perform predictions for a period  between 11 November 2014 and 30 March 2015.
The second tensor is filled with the complete data set and decomposed. 
With the use of the two tensors, the results of the complete dataset can be compared with the results simulated by LSTM. 

\begin{figure}[b!]
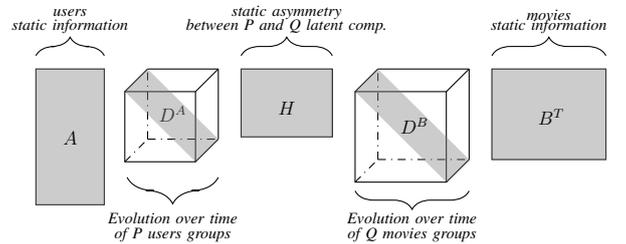

\begin{center}
\includestandalone{tikz_fig/tikz_prtck_movies}
\caption{PARATUCK2 decomposition applied to recommender systems for online video on demand. The model training and predictions are applied on the tensors $\mathscr{D}^A$ and $\mathscr{D}^B$.}
\label{fig::prtck_mov}
\end{center}
\end{figure}

As shown  in the  figure \ref{fig::prtck_mov}, LSTM is trained on $\mathscr{D}^A$ and $\mathscr{D}^B$ for users and movies predictions respectively. 
The matrices $A$, $H$ and $B$ gather static information about the users, the movies and the asymmetry between the P and Q latent factors which have been set to 20 and 30.
Static information defines information which does not evolve over the third dimension, the time. 

\begin{figure}[b!]
\begin{center}
\includegraphics[scale=0.50]{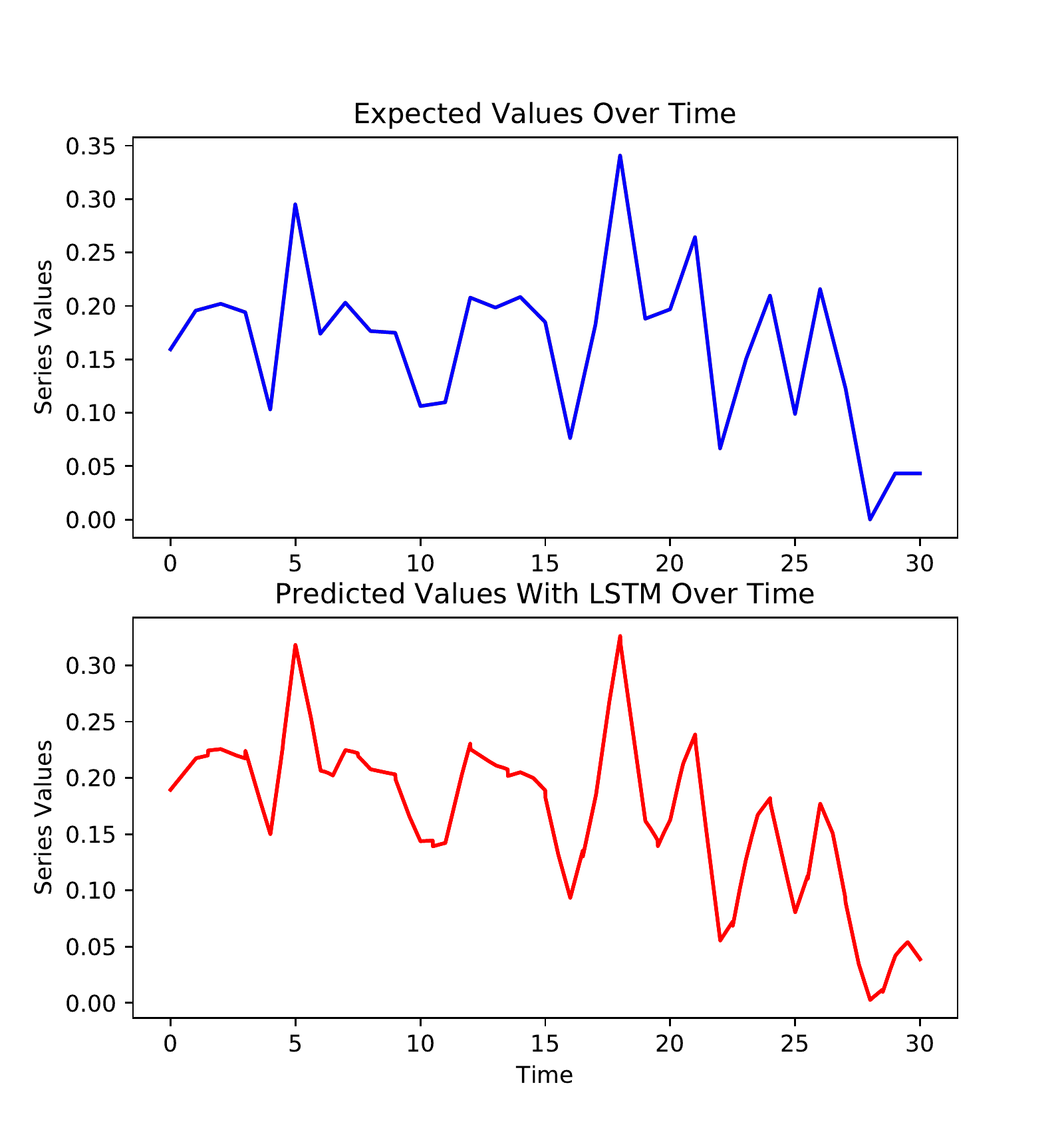}
\caption{PARATUCK2 decomposition applied to recommender systems for online video on demand. The model training and predictions are applied on one latent factor of the tensor $\mathscr{D}^A$.}
\label{fig::2dplot_DA}
\end{center}
\end{figure}

\begin{figure}[b!]
\begin{center}
\includegraphics[scale=0.50]{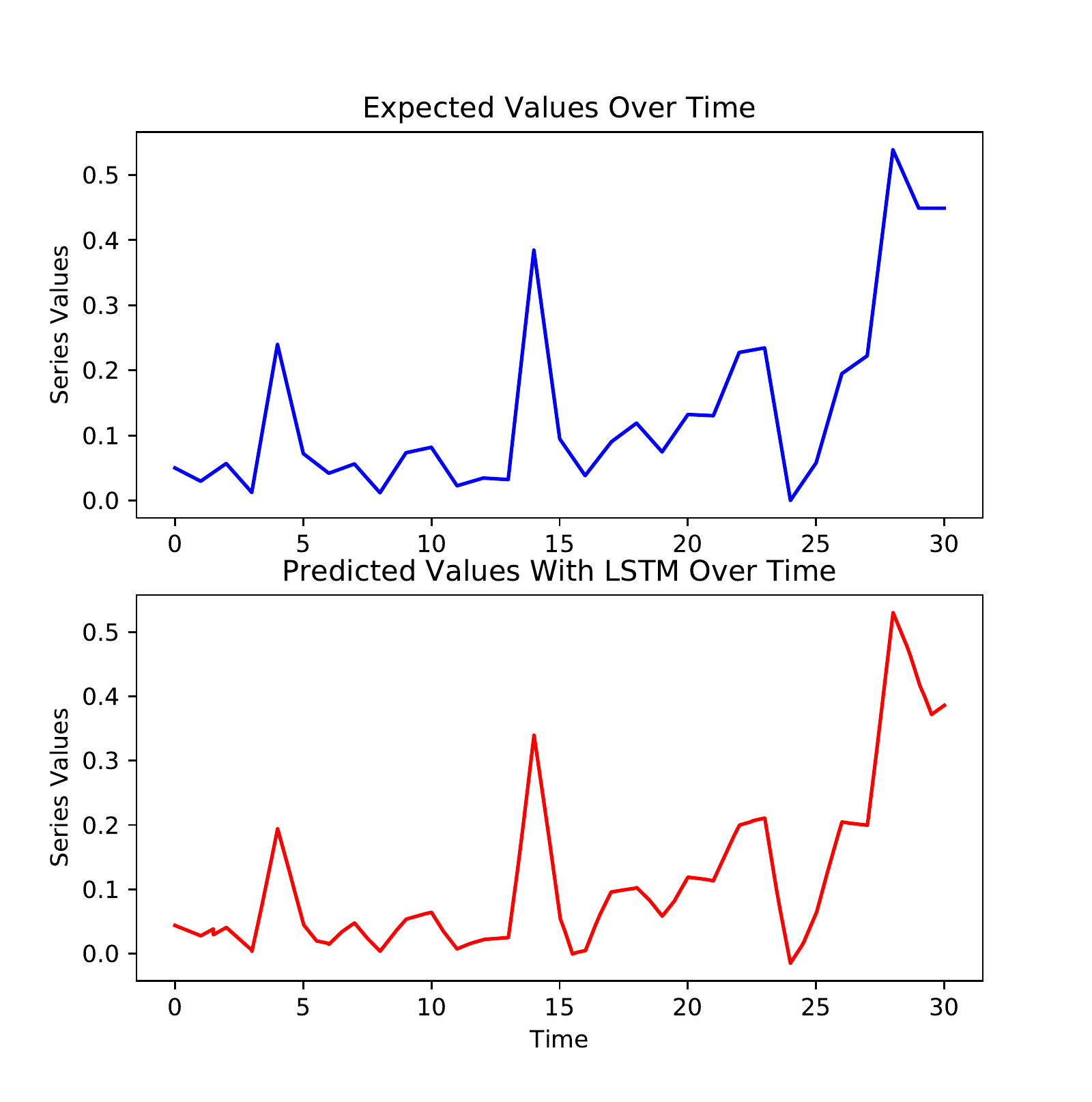}
\caption{PARATUCK2 decomposition applied to recommender systems for online video on demand. The model training and predictions are applied on one latent factor of the tensor $\mathscr{D}^B$.}
\label{fig::2dplot_DB}
\end{center}
\end{figure}

According to figures \ref{fig::2dplot_DA} and \ref{fig::2dplot_DB}, the LSTM predictions seem close to the true values. 
To objectively assess  the accuracy of the predictions, the Mean Absolute Percentage Error (MAE) and the mean directional accuracy (MDA) have been evaluated. 
The MAE is a measure of the prediction of the accuracy for forecasting methods based on the average percentage of difference between the true series and the predicted series. 
The MDA is a measure of the predictions of the accuracy based on the sign of the evolution of the time series. 
It is the average of similar signs in the evolution of the time series between the true values and the predicted values. 

For two random selections of one component of P and one component of Q, respectively related to $\mathscr{D}^A$ and $\mathscr{D}^B$, MAE and MDA criteria are satisfying. 
In table \ref{tbl::mov_stats}, MAE is between 8\% and 12\% and MDA is between 78\% and 80\%.
In addition, it means that the PARATUCK2-LSTM scheme could be used to observe and predict future evolution of P users clusters and Q movies clusters. 

To conclude, this small example illustrates the concepts underlying a  PARATUCK2-LSTM analysis. 
A secondary outcome consists in very good experimental results for series predictions based on the latent factors. These factors, issued through the the tensor decomposition, can be assimilated to different ranks or clusters.
This leads naturally to an applied approach for observing the evolution over time of each latent factors.
Furthermore, the recommender engine could target the recommendation of future movies ranks to future users ranks based on the evolution of $\mathscr{D}^m$ with $m=\left\lbrace A, B \right\rbrace$.

\begin{table}[b!]
\begin{center}
\caption{Statistical tests to assess the accuracy of the predictions applied to recommender systems}
\label{tbl::mov_stats}
\scalebox{1.1}{
\begin{tabular}{|c|c|c|}
\hline
Test & Values for $\mathscr{D}^A$ & Values for $\mathscr{D}^B$\\
\hline
MAE & 0.1182 & 0.0846\\
MDA & 0.8012 & 0.7842\\
\hline
\end{tabular}}
\end{center}
\end{table}

%% file: tikz_fig/tikz_prtck_movies.tex
\begin{tikzpicture}[scale=0.6, transform shape]
\draw  (-4.5,2) rectangle (-3,-1);
\fill [gray, opacity=.4]  (-4.5,2) rectangle (-3,-1);
\draw (-3.75,0.75) node[below]{\large{$A$}};

\draw (2.5,-0.5) rectangle (4.5,1.5) node (v1) {};
\draw (4.5,1.5) -- (5,2);
\draw (2.5,1.5) -- (3,2);
\draw (4.5,-0.5) -- (5,0);
\draw (3,2) -- (5,2);
\draw (5,0) -- (5,2);
\draw [dashdotted] (3,0) -- (5,0);
\draw [dashdotted] (3,0) -- (3,2);
\draw [dashdotted] (2.5,-0.5) -- (3,0);
\fill [gray, opacity=.4] (4.5,-0.5) -- (5,0) -- (3,2) -- (2.5,1.5)  -- cycle;
\draw (-1.5,1.3) node[below]{\large{$D^A$}};

\draw (0,2) rectangle (2,0.5);
\fill [gray, opacity=.4] (0,2) rectangle (2,0.5);
\draw (1,1.4) node[below]{\large{$H$}};

\draw (-2.55,-0.05) rectangle (-1,1.5) node (v1) {};
\draw (-1,1.5) -- (-0.5,2);
\draw (-2.55,1.5) -- (-2.05,2);
\draw (-1,-0.05) -- (-0.5,0.45);
\draw (-2.05,2) -- (-0.5,2);
\draw (-0.5,0.45) -- (-0.5,2);
\draw [dashdotted] (-2.05,0.45) -- (-0.5,0.45);
\draw [dashdotted] (-2.05,0.45) -- (-2.05,2);
\draw [dashdotted] (-2.55,-0.05) -- (-2.05,0.45);
\fill [gray, opacity=.4] (-1,-0.05) -- (-0.5,0.45) -- (-2.05,2.0) -- (-2.55,1.5)  -- cycle;
\draw (3.8,1.05) node[below]{\large{$D^B$}};

\draw  (5.5,2) rectangle (8,0);
\fill [gray, opacity=.4] (5.5,2) rectangle (8,0);
\draw (6.8,1.25) node[below]{\large{$B^T$}};

\draw[decoration={brace,raise=5pt, amplitude=7pt},decorate] (-4.5,2.0) -- node[below=-35pt] {\textit{static information}} (-3,2.0);
\node at (-3.75,3.25) {\textit{users}};

\draw[decoration={brace,mirror,raise=5pt, amplitude=10pt},decorate] (-2.5,-0.10) -- (-0.5,-0.10);
\node at (-1.5,-1.3) {\textit{Evolution over time}};
\node at (-1.5,-1.65) {\textit{of P users groups}};

\draw[decoration={brace,raise=5pt, amplitude=7pt},decorate] (0,2.0) -- node[below=-35pt] {\textit{between P and Q latent comp.}} (2,2.0);
\node at (1,3.25) {\textit{static asymmetry}};

\draw[decoration={brace,mirror,raise=5pt, amplitude=7pt},decorate] (2.5,-0.40) -- (5,-0.40);
\node at (3.75,-1.3) {\textit{Evolution over time}};
\node at (3.75,-1.65) {\textit{of Q movies groups}};

\draw[decoration={brace,raise=5pt, amplitude=7pt},decorate] (5.5,2.0) -- node[below=-35pt] {\textit{static information}} (8,2.0);
\node at (6.75,3.2) {\textit{movies}};

\end{tikzpicture}

%% file: Experiments/SmartContracts/Sim_SmartCntrct.tex
\subsection{Smart Contracts profiling using PARATUCK2-LSTM}
Hyperledger and Ethereum are the two main blockchains for the use of smart contracts. 
The data from Ethereum was collected starting 1 January 2016 and ending 1 July 2016. 
Through the collection process, different data types have been stored, such as the hash key, the sender accounts, the receiver accounts or the blockheights. 
For the considered six months period, 5.5 millions of transactions have been made. This accounts for an average of  26 transactions per sender accounts and 18 transactions per receiver accounts.

As LSTM is trained on historical data, it is well suited to reproduce events that already happened but can hardly simulate events that never happened. For such tasks,  stochastic process models are better choices. 
In the data set, some smart contracts only relate to one transaction, payment or reception. 
Such behavior is difficult to predict and should be considered more like unexpected behavior. For fraud detection, LSTM could be thus appropriate. 
Since our aim  is to predict future interactions based on exchanges that already happened in the past and not to detect suspicious behavior, only the 1\% most active smart contracts have been kept in the training set due to their regular activities.

The complete dataset is divided into two datasets, one collecting events between the  1 January 2016 and the 1 April 2016 and another one between 2 April 2016 and 1 July 2016.
The split of the dataset allows to cross-validate the results between the predictions of exchanges performed by LSTM and the true exchanges performed between the smart contracts.

\begin{figure}[b!]
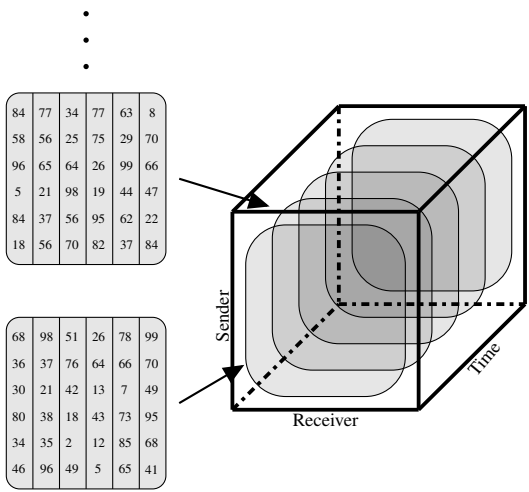

\begin{center}
\includestandalone{tikz_fig/tikz_tnsrCmplt}
\caption{Three$-$way tensor containing Ether amount exchanged between different sender and receiver accounts}
\label{fig::tnsr_fill}
\end{center}
\end{figure}

Following the methodology described in the first experiment for RS, two tensors denoted by $\mathscr{X}\in\mathbb{R}^{I\times J \times K}$ are built from the Ethereum data. 
The first dimension, $I$, lists the sender accounts, the second dimension $J$, the receiver accounts and the third dimension, $K$, the time.
For each tensor, the interaction between a sender and a receiver is represented by the amount of Ether exchanged at a time (figure \ref{fig::tnsr_fill}).
The first tensor is built and decomposed with the first part of the dataset resulting in a tensor size of $\mathbb{R}^{100 \times 200 \times 25}$. 
LSTM is trained over the time period of the first dataset and used to  perform predictions for the period  between 2 April 2016 and 1 July 2016.
The second tensor is filled directly with the complete dataset and decomposed. 
The second tensor has a size of $\mathbb{R}^{100 \times 200 \times 50}$.

\begin{figure}[b!]
\begin{center}
\includestandalone{tikz_fig/tikz_prtck_SC}
\caption{PARATUCK2 decomposition applied to smart contracts profiling. The model training and predictions are applied on the tensors $\mathscr{D}^A$ and $\mathscr{D}^B$.}
\label{fig::prtck_SC}
\end{center}
\end{figure}

\begin{figure}[b!]
\begin{center}
\includegraphics[scale=0.50]{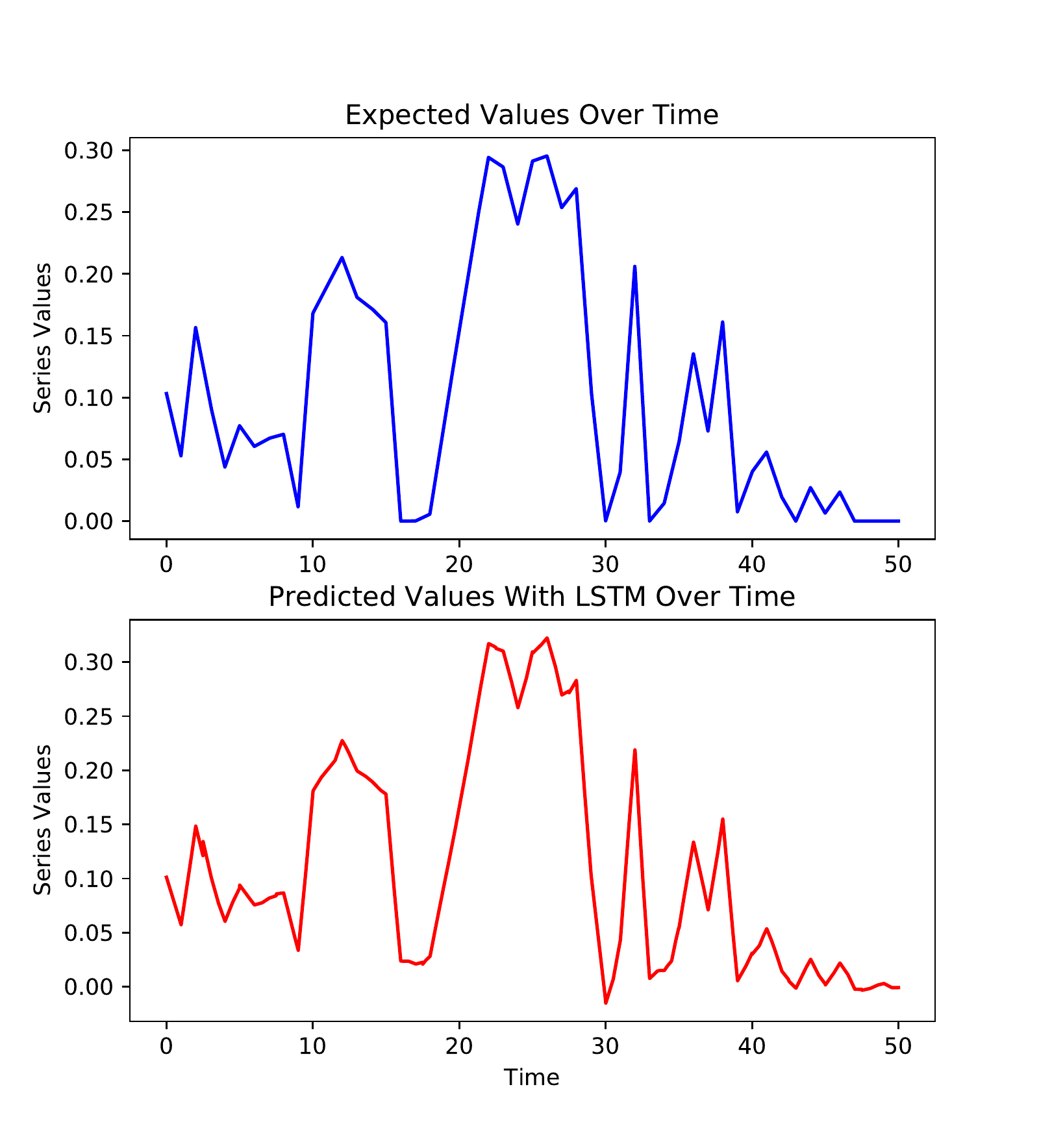}
\caption{PARATUCK2 decomposition applied to smart contracts profiling. The model training and predictions are applied on one latent component of the tensor $\mathscr{D}^A$.}
\label{fig::SC_DA}
\end{center}
\end{figure}

As illustrated in figure \ref{fig::prtck_SC}, the information evolving over time is contained in the tensors $\mathscr{D}^m$ with $m=\left\lbrace A, B \right\rbrace$. 
The matrix $\textbf{A}$ gathers static information regarding $P$ senders groups and the matrix $\textbf{B}$ static information regarding $Q$ receivers groups. 
The matrix $\textbf{H}$ contains the asymmetric information between the $P$ and the $Q$ latent factors which have been set to respectively to 20 and 30.
As a result, the LSTM network is trained on $\mathscr{D}^m$ for the sender and the receiver exchange predictions. 

\begin{table}[b!]
\begin{center}
\caption{Statistical tests to assess the accuracy of the predictions applied to smart contracts}
\label{tbl::SC_stats}
\scalebox{1.1}{
\begin{tabular}{|c|c|c|}
\hline
Test & Values for $\mathscr{D}^A$ & Values for $\mathscr{D}^B$\\
\hline
1 latent var. MAE & 0.1239 & 0.0529\\
1 latent var. MDA & 0.7174 & 0.7166\\
Average MAE & 0.0791 & 0.0776\\
Average MDA & 0.7534 & 0.8073\\
\hline
\end{tabular}}
\end{center}
\end{table}

\begin{figure}[b!]
\begin{center}
\includegraphics[scale=0.50]{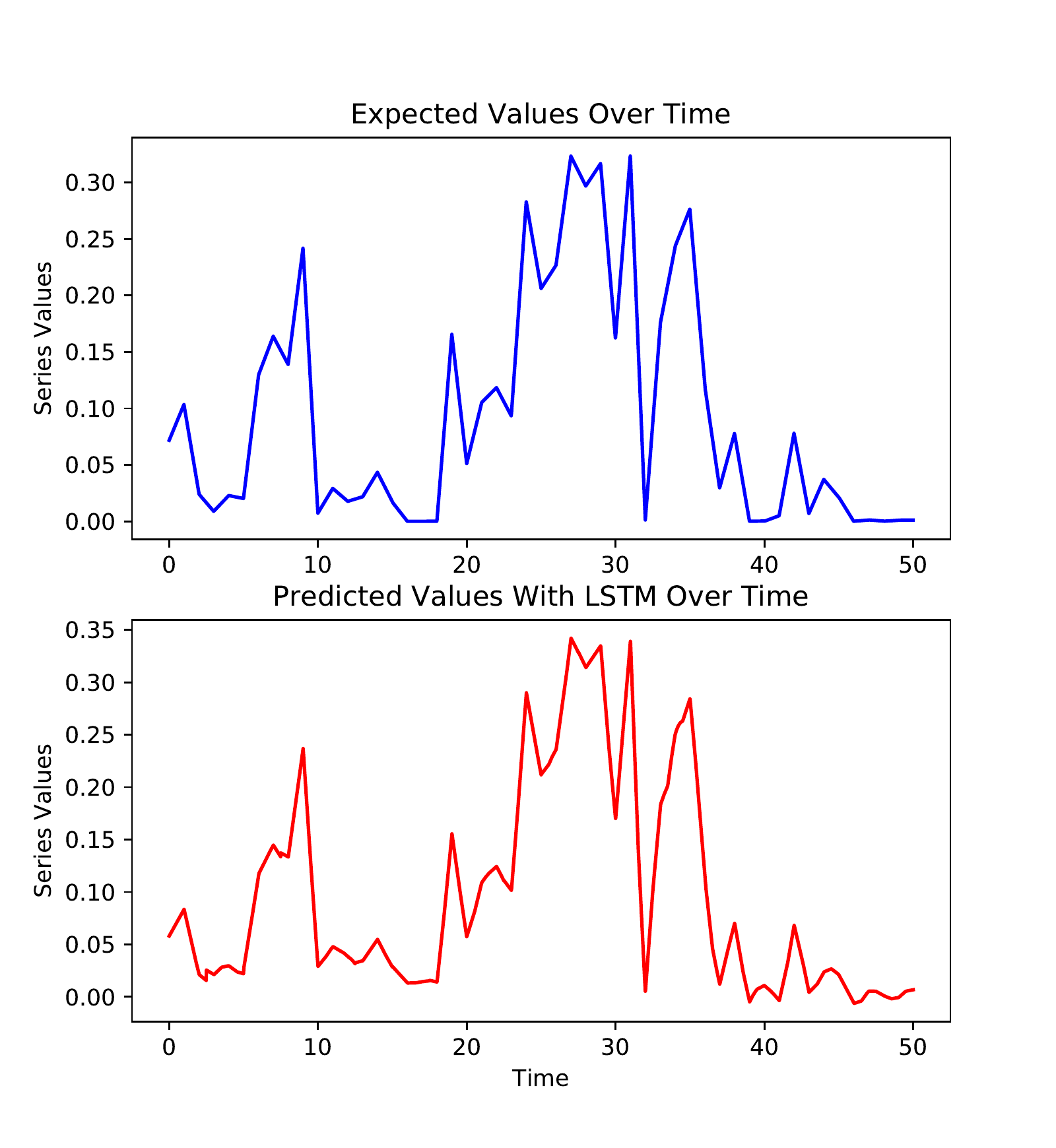}
\caption{PARATUCK2 decomposition applied to smart contracts profiling. The model training and predictions are applied on one latent component of the tensor $\mathscr{D}^B$.}
\label{fig::SC_DB}
\end{center}
\end{figure}

According to figures \ref{fig::SC_DA} and \ref{fig::SC_DB} which show the difference between the true data and the predictions for one rank of the tensors $\mathscr{D}^A$ and $\mathscr{D}^B$, the predictions of Ether exchanges are close to the one observed in the tensor decomposition of the complete true dataset. 
It means that LSTM can be seen as appropriate for the modeling of the smart contracts having regular exchanges.
MAE and MDA have been computed to assess objectively the difference between the simulated results and the true data.

Table \ref{tbl::SC_stats} illustrates the accuracy of the results obtained from the interactions predictions using LSTM.
If we consider only one latent factor selected randomly among $P$ and $Q$, respectively for $\mathscr{D}^A$ and $\mathscr{D}^B$, the MAE is between 5\% and 12\% while the MDA is around 72\%. 
If we consider the average of all latent factors $P$ and $Q$, the MAE is between 78\% and 79\% and the MDA is between 75\% and 81\%.
From statistical point of view, the values for the overall MAE and overall MDA are accurate.
From the highlighted results of this application, the combined approach of PARATUCK2-LSTM delivers good results validated visually and statistically.

%% file: tikz_fig/tikz_tnsrCmplt.tex
\begin{tikzpicture}[scale=0.35, transform shape]
\draw [line width= 0.5mm](0,-0.5) -- (0,7);
\draw [line width= 0.5mm](0,7) -- (7,7);
\draw [line width= 0.5mm](0,-0.5) -- (7,-0.5);
\draw [line width= 0.5mm](7,-0.5) -- (7,7);
\draw [line width= 0.5mm](7,-0.5) -- (11,3.5);
\draw [line width= 0.5mm](11,3.5) -- (11,11);
\draw [line width= 0.5mm](4,11) -- (11,11);
\draw [line width= 0.5mm](0,7) -- (4,11);
\draw [line width= 0.5mm, dashdotted] (4,11) -- (4,3.5);
\draw [line width= 0.5mm, dashdotted] (4,3.5) -- (11,3.5);
\draw [line width= 0.5mm, dashdotted] (0,-0.5) -- (4,3.5);
\draw [line width= 0.5mm](7,7) -- (11,11);

\draw [rounded corners=15pt] (0.5,6.5) rectangle (6.5,0);
\fill [opacity=0.1, rounded corners=15pt] (0.5,6.5) rectangle (6.5,0);

\draw [rounded corners=15pt] (2.5,8.5) rectangle (8.5,2);
\fill [opacity=0.1, rounded corners=15pt] (2.5,8.5) rectangle (8.5,2);

\draw [rounded corners=15pt] (1.5,7.5) rectangle (7.5,1);
\fill [opacity=0.1, rounded corners=15pt] (1.5,7.5) rectangle (7.5,1);

\draw [rounded corners=15pt] (3.5,9.5) rectangle (9.5,3);
\fill [opacity=0.1, rounded corners=15pt] (3.5,9.5) rectangle (9.5,3);

\draw [rounded corners=15pt] (4.5,10.5) rectangle (10.5,4);
\fill [opacity=0.1, rounded corners=15pt] (4.5,10.5) rectangle (10.5,4);

\draw (3.5,-0.5) node[below]{\huge{Receiver}};
\draw (0,3.05) node[above, rotate=90]{\huge{Sender}};
\draw (9.75,1.2) node[above, rotate=50]{\huge{Time}};

\draw (-8.4,2.25) node[right]{\Large{68}};
\draw (-7.35,2.25) node[right]{\Large{98}};
\draw (-6.4,2.25) node[right]{\Large{51}};
\draw (-3.4,2.25) node[right]{\Large{99}};
\draw (-5.4,2.25) node[right]{\Large{26}};
\draw (-4.4,2.25) node[right]{\Large{78}};

\draw (-8.4,1.25) node[right]{\Large{36}};
\draw (-7.35,1.25) node[right]{\Large{37}};
\draw (-6.4,1.25) node[right]{\Large{76}};
\draw (-3.4,1.25) node[right]{\Large{70}};
\draw (-5.4,1.25) node[right]{\Large{64}};
\draw (-4.4,1.25) node[right]{\Large{66}};

\draw (-8.4,0.25) node[right]{\Large{30}};
\draw (-7.35,0.25) node[right]{\Large{21}};
\draw (-6.4,0.25) node[right]{\Large{42}};
\draw (-3.4,0.25) node[right]{\Large{49}};
\draw (-5.4,0.25) node[right]{\Large{13}};
\draw (-4.3,0.25) node[right]{\Large{7}};

\draw (-8.4,-0.75) node[right]{\Large{80}};
\draw (-7.35,-0.75) node[right]{\Large{38}};
\draw (-6.4,-0.75) node[right]{\Large{18}};
\draw (-3.4,-0.75) node[right]{\Large{95}};
\draw (-5.4,-0.75) node[right]{\Large{43}};
\draw (-4.4,-0.75) node[right]{\Large{73}};

\draw (-8.4,-1.75) node[right]{\Large{34}};
\draw (-7.35,-1.75) node[right]{\Large{35}};
\draw (-6.4,-1.75) node[right]{\Large{2}};
\draw (-3.4,-1.75) node[right]{\Large{68}};
\draw (-5.4,-1.75) node[right]{\Large{12}};
\draw (-4.4,-1.75) node[right]{\Large{85}};

\draw (-8.4,-2.75) node[right]{\Large{46}};
\draw (-7.35,-2.75) node[right]{\Large{96}};
\draw (-6.4,-2.75) node[right]{\Large{49}};
\draw (-3.4,-2.75) node[right]{\Large{41}};
\draw (-5.3,-2.75) node[right]{\Large{5}};
\draw (-4.4,-2.75) node[right]{\Large{65}};

\draw [rounded corners=5pt] (-8.5,3) rectangle (-2.5,-3.5);
\fill [opacity=0.1, rounded corners=5pt] (-8.5,3) rectangle (-2.5,-3.5);
\draw (-7.5,3) -- (-7.5,-3.5);
\draw (-6.5,3) -- (-6.5,-3.5);
\draw (-5.5,3) -- (-5.5,-3.5);
\draw (-4.5,3) -- (-4.5,-3.5);
\draw (-3.5,3) -- (-3.5,-3.5);

\draw (-8.4,10.75) node[right]{\Large{84}};
\draw (-7.4,10.75) node[right]{\Large{77}};
\draw (-6.4,10.75) node[right]{\Large{34}};
\draw (-3.25,10.75) node[right]{\Large{8}};
\draw (-5.4,10.75) node[right]{\Large{77}};
\draw (-4.35,10.75) node[right]{\Large{63}};

\draw (-8.4,9.75) node[right]{\Large{58}};
\draw (-7.4,9.75) node[right]{\Large{56}};
\draw (-6.4,9.75) node[right]{\Large{25}};
\draw (-3.4,9.75) node[right]{\Large{70}};
\draw (-5.4,9.75) node[right]{\Large{75}};
\draw (-4.35,9.75) node[right]{\Large{29}};

\draw (-8.4,8.75) node[right]{\Large{96}};
\draw (-7.4,8.75) node[right]{\Large{65}};
\draw (-6.4,8.75) node[right]{\Large{64}};
\draw (-3.4,8.75) node[right]{\Large{66}};
\draw (-5.4,8.75) node[right]{\Large{26}};
\draw (-4.35,8.75) node[right]{\Large{99}};

\draw (-8.3,7.75) node[right]{\Large{5}};
\draw (-7.4,7.75) node[right]{\Large{21}};
\draw (-6.4,7.75) node[right]{\Large{98}};
\draw (-3.4,7.75) node[right]{\Large{47}};
\draw (-5.4,7.75) node[right]{\Large{19}};

\draw (-4.35,7.75) node[right]{\Large{44}};

\draw (-8.4,6.75) node[right]{\Large{84}};
\draw (-7.4,6.75) node[right]{\Large{37}};
\draw (-6.4,6.75) node[right]{\Large{56}};
\draw (-3.4,6.75) node[right]{\Large{22}};
\draw (-5.4,6.75) node[right]{\Large{95}};
\draw (-4.35,6.75) node[right]{\Large{62}};

\draw (-8.4,5.75) node[right]{\Large{18}};
\draw (-7.4,5.75) node[right]{\Large{56}};
\draw (-6.4,5.75) node[right]{\Large{70}};
\draw (-3.4,5.75) node[right]{\Large{84}};
\draw (-5.4,5.75) node[right]{\Large{82}};
\draw (-4.35,5.75) node[right]{\Large{37}};

\draw [rounded corners=5pt] (-8.5,11.5) rectangle (-2.5,5);
\fill [opacity=0.1, rounded corners=5pt] (-8.5,11.5) rectangle (-2.5,5);
\draw (-7.5,11.5) -- (-7.5,5);
\draw (-6.5,11.5) -- (-6.5,5);
\draw (-5.5,11.5) -- (-5.5,5);
\draw (-4.5,11.5) -- (-4.5,5);
\draw (-3.5,11.5) -- (-3.5,5);

\fill  (-5.5,12.5) circle (0.10);
\fill  (-5.5,13.5) circle (0.10);
\fill  (-5.5,14.5) circle (0.10);

\draw [-triangle 45, thick=0.25mm] (-2,8.25) -- (1.4,7.2);
\draw [-triangle 45, thick=0.25mm] (-2,-0.25) -- (0.45,1.2);
\end{tikzpicture}

%% file: tikz_fig/tikz_prtck_SC.tex
\begin{tikzpicture}[scale=0.6, transform shape]
\draw  (-4.5,2) rectangle (-3,-1);
\fill [gray, opacity=.4]  (-4.5,2) rectangle (-3,-1);
\draw (-3.75,0.75) node[below]{\large{$A$}};

\draw (2.5,-0.5) rectangle (4.5,1.5) node (v1) {};
\draw (4.5,1.5) -- (5,2);
\draw (2.5,1.5) -- (3,2);
\draw (4.5,-0.5) -- (5,0);
\draw (3,2) -- (5,2);
\draw (5,0) -- (5,2);
\draw [dashdotted] (3,0) -- (5,0);
\draw [dashdotted] (3,0) -- (3,2);
\draw [dashdotted] (2.5,-0.5) -- (3,0);
\fill [gray, opacity=.4] (4.5,-0.5) -- (5,0) -- (3,2) -- (2.5,1.5)  -- cycle;
\draw (-1.5,1.3) node[below]{\large{$D^A$}};

\draw (0,2) rectangle (2,0.5);
\fill [gray, opacity=.4] (0,2) rectangle (2,0.5);
\draw (1,1.4) node[below]{\large{$H$}};

\draw (-2.55,-0.05) rectangle (-1,1.5) node (v1) {};
\draw (-1,1.5) -- (-0.5,2);
\draw (-2.55,1.5) -- (-2.05,2);
\draw (-1,-0.05) -- (-0.5,0.45);
\draw (-2.05,2) -- (-0.5,2);
\draw (-0.5,0.45) -- (-0.5,2);
\draw [dashdotted] (-2.05,0.45) -- (-0.5,0.45);
\draw [dashdotted] (-2.05,0.45) -- (-2.05,2);
\draw [dashdotted] (-2.55,-0.05) -- (-2.05,0.45);
\fill [gray, opacity=.4] (-1,-0.05) -- (-0.5,0.45) -- (-2.05,2.0) -- (-2.55,1.5)  -- cycle;
\draw (3.8,1.05) node[below]{\large{$D^B$}};

\draw  (5.5,2) rectangle (8,0);
\fill [gray, opacity=.4] (5.5,2) rectangle (8,0);
\draw (6.8,1.25) node[below]{\large{$B^T$}};

\draw[decoration={brace,raise=5pt, amplitude=7pt},decorate] (-4.5,2.0) -- node[below=-35pt] {\textit{static information}} (-3,2.0);
\node at (-3.75,3.25) {\textit{senders}};

\draw[decoration={brace,mirror,raise=5pt, amplitude=10pt},decorate] (-2.5,-0.10) -- (-0.5,-0.10);
\node at (-1.5,-1.3) {\textit{Evolution over time}};
\node at (-1.5,-1.65) {\textit{of P sender groups}};

\draw[decoration={brace,raise=5pt, amplitude=7pt},decorate] (0,2.0) -- node[below=-35pt] {\textit{between P and Q latent comp.}} (2,2.0);
\node at (1,3.25) {\textit{static asymmetry}};

\draw[decoration={brace,mirror,raise=5pt, amplitude=7pt},decorate] (2.5,-0.40) -- (5,-0.40);
\node at (3.75,-1.3) {\textit{Evolution over time}};
\node at (3.75,-1.65) {\textit{of Q receivers groups}};

\draw[decoration={brace,raise=5pt, amplitude=7pt},decorate] (5.5,2.0) -- node[below=-35pt] {\textit{static information}} (8,2.0);
\node at (6.75,3.2) {\textit{receivers}};

\end{tikzpicture}